\documentclass[twocolumn,trackchanges,times,twocolappendix]{aastex63}
\usepackage{graphicx}
\usepackage{amssymb}
\usepackage{latexsym}
\usepackage{longtable}
\usepackage{epsf}
\usepackage{color}
\usepackage{ulem}
\usepackage{here}

\received{February 11, 2020}
\revised{May 10, 2020}
\accepted{May 17, 2020}
\submitjournal{Astrophysical Journal}
\shorttitle{Element Stratification in SNR G344.7--0.1}

\begin{document}

\title{Element Stratification in the Middle-aged SN I\lowercase{a} Remnant G344.7--0.1}

\author{Kotaro Fukushima}
\affil{Department of Physics, Tokyo University of Science, 1-3 Kagurazaka, Shinjuku-ku, Tokyo 162-8601, Japan}
\author{Hiroya Yamaguchi}
\affil{Institute of Space and Astronautical Science, JAXA, 3-1-1 Yoshinodai, Sagamihara, Kanagawa 229-8510, Japan}
\affil{Department of Physics, The University of Tokyo, 7-3-1 Hongo, Bunkyo-ku, Tokyo 113-0033, Japan}
\author{Patrick O. Slane}
\affil{Harvard-Smithsonian Center for Astrophysics, 60 Garden Street, Cambridge, MA 02138, USA}
\author{Sangwook Park}
\affil{Box 19059, Department of Physics, University of Texas at Arlington, Arlington, TX 76019, USA}
\author{Satoru Katsuda}
\affil{Graduate School of Science and Engineering, Saitama University, 255 Shimo-Ohkubo, Sakura, Saitama 338-8570, Japan}
\author{Hidetoshi Sano}
\affil{Institute for Advanced Research, Nagoya University, Furo-cho, Chikusa-ku, Nagoya 464-8601, Japan}
\affil{Department of Physics, Nagoya University, Furo-cho, Chikusa-ku, Nagoya 464-8601, Japan}
\author{Laura A. Lopez}
\affil{Department of Astronomy, The Ohio State University, 140 W. 18th Avenue, Columbus, Ohio 43210, USA}
\affil{Center for Cosmology and AstroParticle Physics, The Ohio State University, 191 W. Woodruff Avenue, Columbus, OH 43210, USA}
\affil{Niels Bohr Institute, University of Copenhagen, Blegdamsvej 17, DK-2100 Copenhagen, Denmark}
\author{Paul P. Plucinsky}
\affil{Harvard-Smithsonian Center for Astrophysics, 60 Garden Street, Cambridge, MA 02138, USA}
\author{Shogo B. Kobayashi}
\affil{Department of Physics, Tokyo University of Science, 1-3 Kagurazaka, Shinjuku-ku, Tokyo 162-8601, Japan}
\author{Kyoko Matsushita}
\affil{Department of Physics, Tokyo University of Science, 1-3 Kagurazaka, Shinjuku-ku, Tokyo 162-8601, Japan}

\shortauthors{Fukushima et al.}

\begin{abstract}

Despite their importance, a detailed understanding of Type Ia supernovae (SNe Ia) remains elusive.
X-ray measurements of the element distributions in supernova remnants (SNRs) offer
important clues for understanding the explosion and nucleosynthesis mechanisms for SNe Ia.
However, it is challenging to observe the entire ejecta mass in X-rays for young SNRs,
because the central ejecta may not have been heated by the reverse shock yet.
Here we present over 200 kilosecond Chandra observations of the Type Ia SNR G344.7--0.1,
whose age is old enough for the reverse shock to have reached the SNR center,
providing an opportunity to investigate the distribution of the entire ejecta mass.
We reveal a clear stratification of heavy elements with a centrally peaked distribution of the Fe ejecta
surrounded by intermediate-mass elements (IMEs: Si, S, Ar Ca) with an arc-like structure.
The centroid energy of the Fe K emission is marginally lower in the central Fe-rich region than in
the outer IME-rich regions, suggesting that the Fe ejecta were shock-heated more recently.
These results are consistent with the prediction for standard SN Ia models, where the heavier elements
are synthesized in the interior of an exploding white dwarf. We find, however, that the peak location
of the Fe K emission is slightly offset to the west with respect to the geometric center of the SNR.
This apparent asymmetry is likely due to the inhomogeneous density distribution of the ambient
medium, consistent with our radio observations of the ambient molecular and neutral gas.

\end{abstract}


\section{Introduction}
\label{sec:intro}

\begin{deluxetable*}{ccCCC}[ht!]
\tablecaption{Observation Log \label{tab:observation}}
\tablenum{1}
\tablewidth{0pt}
\tablehead{
\colhead{Obs.\ ID} & \colhead{Date} & \colhead{Exposure Time($\mathrm{ks}$)} & \colhead{Aim Point\tablenotemark{a}} & \colhead{Roll Angle}}
\startdata
20308 & 2018 May 16 & 29.2 & 255^\circ.99, -41^\circ.72 & 50^\circ.21\\
20309 & 2018 May 12 & 55.3 & 255^\circ.99, -41^\circ.72 & 51^\circ.21\\
21093 & 2018 May 18 & 33.6 & 255^\circ.99, -41^\circ.72 & 50^\circ.21\\
21094 & 2018 May 19 & 14.9 & 255^\circ.99, -41^\circ.72 & 50^\circ.21\\
21095 & 2018 May 20 & 26.7 & 255^\circ.99, -41^\circ.72 & 50^\circ.21\\
21096 & 2018 Jul 3 & 27.2 & 256^\circ.06, -41^\circ.71 & 303^\circ.2\\
21117 & 2018 Jul 5 & 18.8 & 256^\circ.06, -41^\circ.71 & 303^\circ.2\\
\enddata
\tablenotetext{a}{In the $\text{(R.A., decl.)}_{\text{J2000}}$ coordinate.}
\end{deluxetable*}

Type Ia supernovae (SNe Ia) are thought to result from thermonuclear explosions of white dwarfs in
a binary system. Since the peak luminosity in the optical band is almost uniform among the
objects, SNe Ia can be utilized as distance indicators in cosmology \citep[e.g.,][]{Riess98,Perlmutter98}.
SNe Ia also play an important role as major suppliers of iron, contributing to the chemical
enrichment of the Milky Way \citep[e.g.,][]{Reddy06}
and galaxy clusters \citep[e.g.,][]{Sato07}.
Despite this importance, however, many fundamental aspects of these explosions remain elusive.

X-ray observations of supernova remnants (SNRs) offer a unique way to address the relevant
open questions, as they allow us to measure the composition and distribution of heavy elements
that were synthesized during the SN explosion.
In fact, recent observational studies of young, luminous Type Ia SNRs, such as Kepler, Tycho,
and SN\,1006, have provided important insights into the evolution of the explosions of their progenitors
\citep[e.g.,][and references therein]{Vink17,Decourchelle17,Katsuda17}.  
In these young historical SNRs, however, the central ejecta might not yet have been heated
by the reverse shock and thus are invisible in X-rays. 
Thus, while some extensive studies have been performed with these young Ia SNRs,
our knowledge of how the SN Ia explosion begins or what happens
in the deepest layers of the exploding white dwarf is still limited.

The Type Ia SNR G344.7--0.1 is an ideal target for a comprehensive X-ray ejecta study of SN Ia,
because its age is significantly older (3--6\,kyr: \citealt{Combi10}; \citealt{Giacani11})
than the historical Ia SNRs, for which the reverse shock has likely reached the SNR center.
This SNR was discovered in the radio band with the Molonglo Observatory Synthesis
Telescope (MOST) and the Parkes Observatory \citep{Clark75}. 
X-rays from this region were first detected by the ASCA Galactic Plane Survey \citep{Sugizaki01}, and 
identified as emission originating from an optically thin thermal plasma associated with the SNR \citep{Yamauchi05}.
The following studies with Chandra and XMM-Newton enabled more detailed imaging and 
spectroscopic analysis, where the abovementioned SNR ionization age was estimated \citep{Combi10,Giacani11}.
G344.7--0.1 was initially classified as a core-collapse (CC) SNR, based on its highly asymmetric
X-ray morphology \citep{Lopez11}
and possible association with star forming regions \citep{Giacani11}.
However, elemental abundances of the SN ejecta were not well constrained in these studies,
mainly due to the short exposure, $\sim$\,25\,ks for both Chandra and XMM-Newton.

The situation was largely changed by Suzaku observations, where strong, ejecta-dominated
Fe K emission was detected \citep{Yamaguchi12b}.
Its centroid energy $\sim$\,6.46\,keV corresponds to a relatively low ionization state (Fe$^{17+}$),
which is proposed to be a common characteristic for Type Ia SNRs \citep{Yamaguchi14b}. 
Notably, the atomic processes involved in the Fe K fluorescence emission are theoretically 
well understood \citep[e.g.,][]{Yamaguchi14a}. 
Thus, the detection of the Fe K emission from this SNR is a remarkable advantage
compared to other Type Ia SNRs (with similar ages of several kyr, 
e.g., G299.2--2.9: \citealt{Post14}; DEM\,L71: \citealt{Hughes03}),
where the X-ray spectra are dominated by Fe L-shell emission with highly uncertain atomic physics.
Unfortunately, the angular resolution of Suzaku was not good enough to constrain the detailed
distribution of the Fe ejecta in G344.7--0.1.
This motivated us to carry out deep Chandra observations to reveal the distribution
of the entire SN ejecta mass, including Fe, the major nucleosynthesis product of SNe Ia.

It should also be noted that the previous Chandra observation detected a point-like source in
the soft X-ray band, named CXOU~J170357.8--414302, at the geometrical center of G344.7--0.1
\citep{Combi10}.
Since the absorption column density of this source ($N_{\rm H} \approx 1 \times 10^{22}$\,cm$^{-2}$)
was significantly lower than that of the SNR ($5 \times 10^{22}$\,cm$^{-2}$),
it was concluded in their work that the source was a foreground object unassociated with the SNR.
However, one may still suspect the possibility of a central compact object (CCO) originating
from a CC progenitor, because of the poor photon statistics from the previous observation.
Our additional 
200 ks Chandra observations also allow us to confirm 
that this object is unrelated to G344.7--0.1 with better statistics
and to search for other point-like sources near the SNR center.

In Section \ref{sec:observation}, we describe details of our Chandra observations and data reduction.
In Section \ref{sec:analysis}, we present our imaging and spectral analysis.
We discuss the results in Section \ref{sec:discuss} and present our conclusions in Section \ref{sec:con}.
The errors quoted in this work are at $1\sigma$ confidence level, 
unless otherwise stated.

\section{Observations and Data Reduction}
\label{sec:observation}

We used the ACIS-I array \citep{Garmire03} 
for our deep observations of G344.7--0.1 to
cover the whole X-ray emitting region ($\sim 8'$ in diameter) within a single field of view (FoV).
The ACIS-I has a lower background level than the S3 chip, especially at energies above 5\,keV,
offering a great advantage for our primary objective: detection and localization of the Fe K emission.

\begin{figure}[ht!]
\gridline{\fig{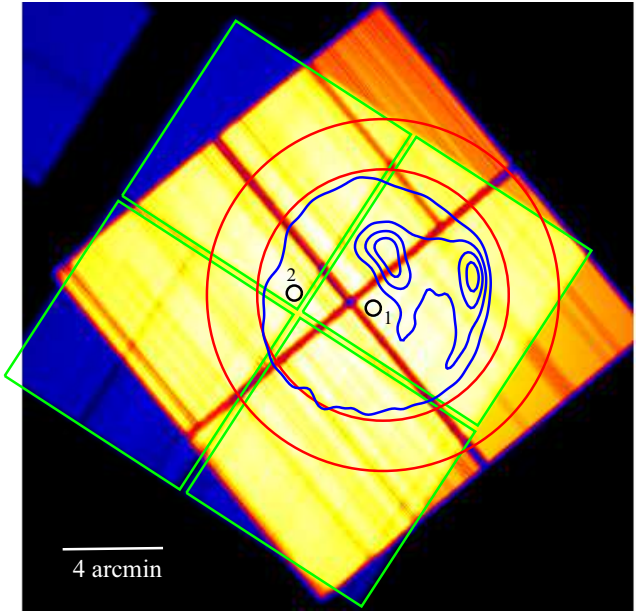}{1.0\columnwidth}{}
	}
\caption{Merged exposure map of our deep {\it Chandra} observations presented in this paper. 
The maps from the two observation series are combined (see the text for the details). 
The green squares indicate the ACIS-I FoV of the shorter ($\sim$\,45\,ks) observation series in 2018 July. 
The black circles labeled 1 and 2 indicate the on-axis aim points of the former and latter observations, 
respectively. The red annulus is the background region used for the spectral analysis.
The blue contours are the $843\ \mathrm{MHz}$ radio image of G344.7--0.1 taken from
the MOST Supernova Remnant Catalog \citep{WG96}. 
North is up and east is to the left. \label{fig:snr}}
\end{figure}

As summarized in Table\,\ref{tab:observation}, the observations were split into two series,
one in May and the other in 2018 July,
with the exposure totaling $\sim$\,205\,ks. Our quick-look analysis after the first observations series 
revealed that the Fe K emission partially fell into the gaps between the ACIS-I CCDs,
where the exposure is reduced as shown in Figure\,\ref{fig:snr}.
Therefore, we slightly shifted the aim point for the later observation series
such that the Fe-rich regions received the maximum exposure.
The difference in the aim points and roll angles between the two series is indicated  
in Figure\,\ref{fig:snr} with the integrated exposure map overplotted with the radio intensity contours of the SNR.

We reprocess the data in accordance with the standard reduction procedures using CIAO version 4.11
\citep{Fruscione06} 
and the calibration database (CALDB) version 4.8.2.
For spectral analysis presented in \S\ref{subsec:ext}, and Appendix \ref{sec:background},
we extract ACIS spectra and generated
RMFs and ARFs for each individual ObsID, and those taken in the same observation series
(i.e., May or July) are summed up to improve the photon statistics. The two merged spectra taken
from the same sky regions but different series are jointly fitted with the identical spectral parameters.

\section{Analysis and Results}
\label{sec:analysis}

\subsection{Imaging Analysis}
\label{subsec:img}

\subsubsection{Extended Emission}

\begin{figure*}[ht!]
\gridline{\fig{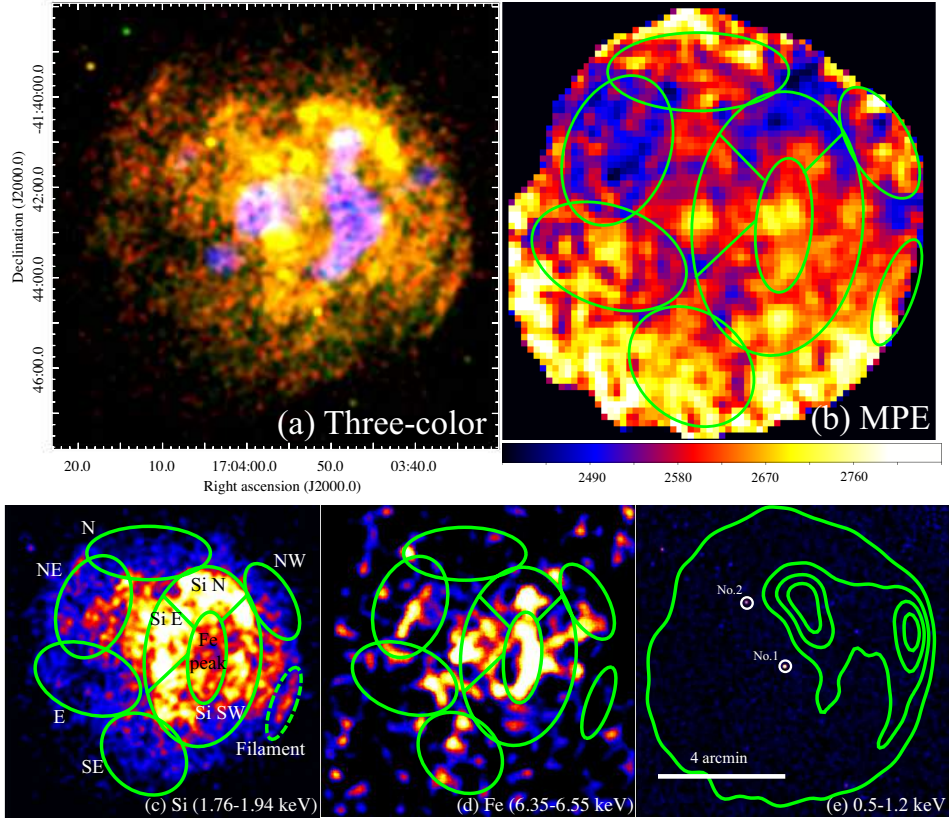}{0.7\textwidth}{}}
\caption{ACIS-I images of G344.7--0.1. \label{fig:band}
(a) Three-color image, where red, green, and blue correspond to emission
from the Si K (1.76--1.94\,keV), hard continuum (3.0--6.0\,keV), 
and Fe K (6.35--6.55\,keV) bands, respectively. \ 
(b) Mean energies of the 1.5--5.0\,keV photons. The color bar is in units of eV.
The spectrum extraction regions are indicated by the green ellipses. \ 
(c) and (d) Narrowband images in the 1.76--1.94\,keV and 6.35--6.55\,keV bands, 
corresponding to the Si K and Fe K emission. \ 
(e) Soft X-ray image in the 0.5--1.2\,keV band. 
The radio image of the SNR is overplotted in contours.
Two bright point-like sources are identified.}
\end{figure*}

\begin{figure}[ht!]
\gridline{\fig{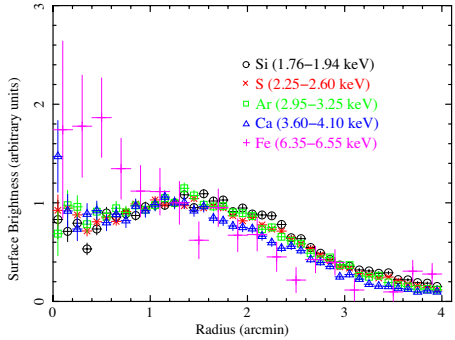}{0.7\columnwidth}{}}
\caption{Radial profiles of the surface brightness in the K-shell emission bands of
Si, S, Ar, Ca, and Fe, where the continuum flux is unsubtracted.
The Fe peak position shown in Figure \ref{fig:band} is taken as the origin of these profiles.
Surface brightness is extracted from the exposure-corrected flux images of the corresponding 
energy bands and then linearly scaled so that the normalization of the bin at $r = 1'.35$ 
becomes unity for each element. \label{fig:surbri}}
\end{figure}

Figure\,\ref{fig:band}(a) shows a three-color, exposure-corrected image of G344.7--0.1, where red, green, 
and blue correspond to emission from the Si K (1.76--1.94\,keV), hard continuum (3.0--6.0\,keV), 
and Fe K (6.35--6.55\,keV) band, respectively.
The image indicates that the softer X-ray emission (red) spreads across the entire SNR, 
whereas the hard X-ray continuum (green) is relatively faint in the northeast region,
implying a temperature gradient from the southwest (high) to the northeast (low). 
This gradient is more clearly highlighted in Figure\,\ref{fig:band}(b), a map of the mean energy of 
the detected photons in the broad energy band of 1.5--5.0\,keV generated using 
the {\tt mean\_energy\_map} script in the {\tt CIAO} package.\footnote{The mean 
photon energy (MPE) is defined as $\langle E(x, y) \rangle$ = $({\sum_i E_i(x, y)})/N(x, y)$
(where $E_i$ and N are the energy of each photon and the total number of photons detected in 
the pixel $x, y$). An MPE map is known to represent the temperature distribution, when the photons are 
extracted from the broad energy band \citep[e.g.,][]{Troja08,David09}.
}
Note, however, that this energy band includes several emission lines,
and thus a more quantitative spectral analysis (\S\ref{subsec:ext}) is required
to determine the accurate temperature distribution. 


Figures\,\ref{fig:band}(c) and (d) are the narrowband images of the Si K and Fe K emission, 
identical to the red and blue images in Figure\,\ref{fig:band}(a) respectively. 
It is remarkable that the Fe K line-enhanced region is surrounded by 
the arc-like structure of the Si K emission. Also notable is that the Fe K peak location 
is about $2'$ offset to the west with respect to the geometric center of the SNR
determined by the radio band image.
In addition to the arc of the Si K emission, a filamentary structure is found at the west rim 
in the same energy band, indicated with the dashed ellipse in Figure\,\ref{fig:band}(c). 
All of these features are discovered in this work for the first time, 
owing to the deep observations with the Chandra ACIS-I.

We generate in Figure\,\ref{fig:surbri} radial profiles of the surface brightness in the narrow energy bands
that correspond to the K-shell emission of Si (1.76--1.94\,keV),
S (2.25--2.60\,keV), Ar (2.95--3.25\,keV), Ca (3.60--4.10\,keV), and Fe (6.35--6.55\,keV).
We confirm that the Fe K emission profile indeed peaks at the interior to other species. 
We also find that the profiles of the intermediate-mass elements (IMEs) are similar to each other.


\subsubsection{Detected Point-like Sources}

\begin{deluxetable*}{cCCCC|cCCCC}[ht!]
\tablecaption{List of the Detected Point-like Sources \label{tab:srclist}}
\tablenum{2}
\tablewidth{0pt}
\tablehead{
\colhead{No.} &
\colhead{$\mathrm{(R.A. , decl)_{J2000}}$} &
\colhead{\tt PSFRATIO\tablenotemark{a}} &
\colhead{Soft\tablenotemark{b}} &
\colhead{Hard\tablenotemark{c}} &
\colhead{No.} &
\colhead{$\mathrm{(R.A. , decl)_{J2000}}$} &
\colhead{\tt PSFRATIO\tablenotemark{a}} &
\colhead{Soft\tablenotemark{b}} &
\colhead{Hard\tablenotemark{c}}
}
\startdata
1 & 17:03:57.8,-41:43:02.2 & 0.56 & 146 & 44 & 26 & 17:03:36.8,-41:40:46.5 & 0.35 & 29 & 65 \\
2 & 17:04:04.3,-41:41:01.0 & 0.48 & 114 & 38 & 27 & 17:03:54.4,-41:40:19.9 & 0.70 & 31 & 59 \\
3 & 17:03:47.5,-41:46:47.1 & 0.77 & 65 & 67 & 28 & 17:04:09.3,-41:40:11.3 & 0.69 & 22 & 29 \\
4 & 17:04:04.1,-41:45:36.2 & 0.24 & 24 & 2 & 29 & 17:04:19.3,-41:38:56.3 & 0.84 & 29 & 85 \\
5 & 17:04:01.2,-41:43:35.6 & 0.25 & 8 & 2 & 30 & 17:03:55.1,-41:46:33.7 & 0.90 & 56 & 110 \\
6 & 17:03:53.4,-41:42:32.2 & 0.34 & 35 & 27 & 31 & 17:03:53.9,-41:41:33.3 & 0.62 & 36 & 80 \\
7 & 17:03:45.7,-41:41:56.4 & 0.43 & 46 & 33 & 32 & 17:03:49.8,-41:41:01.8 & 0.81 & 78 & 128 \\
8 & 17:04:09.1,-41:39:56.9 & 0.64 & 28 & 31 & 33 & 17:03:38.5,-41:45:15.2 & 0.64 & 41 & 124 \\
9 & 17:04:07.0,-41:39:49.6 & 0.72 & 32 & 17 & 34 & 17:03:37.2,-41:44:36.8 & 0.80 & 137 & 306 \\
10 & 17:04:11.8,-41:39:31.0 & 0.33 & 17 & 1 & 35 & 17:03:52.7,-41:44:27.8 & 0.51 & 20 & 32 \\
11 & 17:03:41.6,-41:46:28.9 & 0.47 & 21 & 39 & 36 & 17:03:35.9,-41:44:22.0 & 0.51 & 49 & 118 \\
12 & 17:04:03.0,-41:44:06.2 & 0.28 & 8 & 3 & 37 & 17:03:46.8,-41:44:13.8 & 0.68 & 39 & 109 \\
13 & 17:04:11.7,-41:39:53.4 & 0.28 & 11 & 4 & 38 & 17:03:44.7,-41:43:57.6 & 0.37 & 15 & 34 \\
14 & 17:04:20.1,-41:39:47.7 & 0.35 & 28 & 13 & 39 & 17:03:34.4,-41:43:20.2 & 0.67 & 80 & 170 \\
15 & 17:04:18.5,-41:39:21.0 & 0.38 & 46& 22 & 40 & 17:03:55.2,-41:43:12.4 & 0.30 & 7 & 7 \\
16 & 17:04:16.3,-41:38:37.9 & 0.36 & 24 & 7 & 41 & 17:03:56.1,-41:43:00.6 & 0.38 & 10 & 20 \\
17 & 17:04:14.4,-41:38:34.2 & 0.62 & 80 & 70 & 42 & 17:03:43.9,-41:42:43.3 & 0.48 & 25 & 57 \\
18 & 17:03:38.8,-41:46:36.1 & 0.45 & 40 & 38 & 43 & 17:03:37.0,-41:42:20.0 & 0.76 & 116 & 191 \\
19 & 17:03:36.8,-41:39:40.1 & 0.55 & 35 & 44 & 44 & 17:04:00.3,-41:42:07.1 & 0.51 & 8 & 14 \\
20 & 17:03:51.7,-41:44:45.8 & 0.95 & 118 & 187 & 45 & 17:03:46.1,-41:40:52.3 & 0.24 & 14 & 20 \\
21 & 17:04:04.7,-41:43:47.9 & 0.39 & 15 & 9 & 46 & 17:03:43.4,-41:40:55.5 & 0.60 & 70 & 130 \\
22 & 17:04:02.3,-41:46:23.0 & 0.60 & 39 & 39 & 47 & 17:03:50.4,-41:40:16.8 & 0.90 & 105 & 160 \\
23 & 17:04:19.7,-41:42:52.4 & 0.43 & 19 & 10 & 48 & 17:04:11.1,-41:40:01.8 & 0.91 & 51 & 69 \\
24 & 17:03:33.9,-41:45:49.8 & 0.60 & 68 & 108 & 49 & 17:03:53.4,-41:38:37.4 & 0.91 & 41 & 130 \\
25 & 17:04:14.0,-41:41:28.7 & 0.79 & 22 & 31 &&&&&\\
\enddata
\tablenotetext{a}{A parameter of {\tt wavdetect} that indicates the spatial extension of the sources 
	with respect to the PSF size. The source can be regarded as a true point source when this value is 
	the unity or less.}
\tablenotetext{b}{Photon counts in 0.5--2.0\,keV.}
\tablenotetext{c}{Photon counts in 2.0--9.0\,keV.}
\end{deluxetable*}

Figure\,\ref{fig:band}(e) shows a 0.5--1.2\,keV image of the same region as the other panels of Figure\,\ref{fig:band}. 
Because of the large foreground extinction ($N_{\rm H} \sim 5 \times 10^{22}$\,cm$^{-2}$: \citealt{Yamauchi05,Combi10,Giacani11,Yamaguchi12b}), 
soft X-rays from G344.7--0.1 are almost fully absorbed. 
Therefore, photons in this energy band are likely dominated by foreground objects. 
Our simple eyeball inspection detects the two bright point-like sources, labeled No.\,1 and 2 in Figure\,\ref{fig:band}(e). 
Source No.\,1 corresponds to CXOU~J170357.8--414302, previously reported by \cite{Combi10}. 
This source is located near the geometric center of the SNR,
and does not positionally coincide with the Fe K peak location.
The other (No.\,2) is newly detected with our deep Chandra observation. 
We also search for fainter point sources using the {\tt wavdetect} algorithm in the two energy 
ranges of 0.5--1.2\,keV and 1.2--2.5\,keV with the Encircled Counts Fraction ({\tt ECF})
parameter set to 0.9. 
Table\,\ref{tab:srclist} gives the detected sources with their coordinates, spatial extent relative to the size of 
the point spread function ({\tt PSFRATIO}), and the counts in the 0.5--2.0\,keV (soft) and 
2.0--9.0\,keV (hard) bands. We regard those with {\tt PSFRATIO} $<$\,1 as true point sources. 
About 50 sources are detected based on these criteria.


\subsection{Spectral Analysis}
\label{subsec:ext}

\subsubsection{Extended Emission}

\begin{figure*}[ht!]
\gridline{\fig{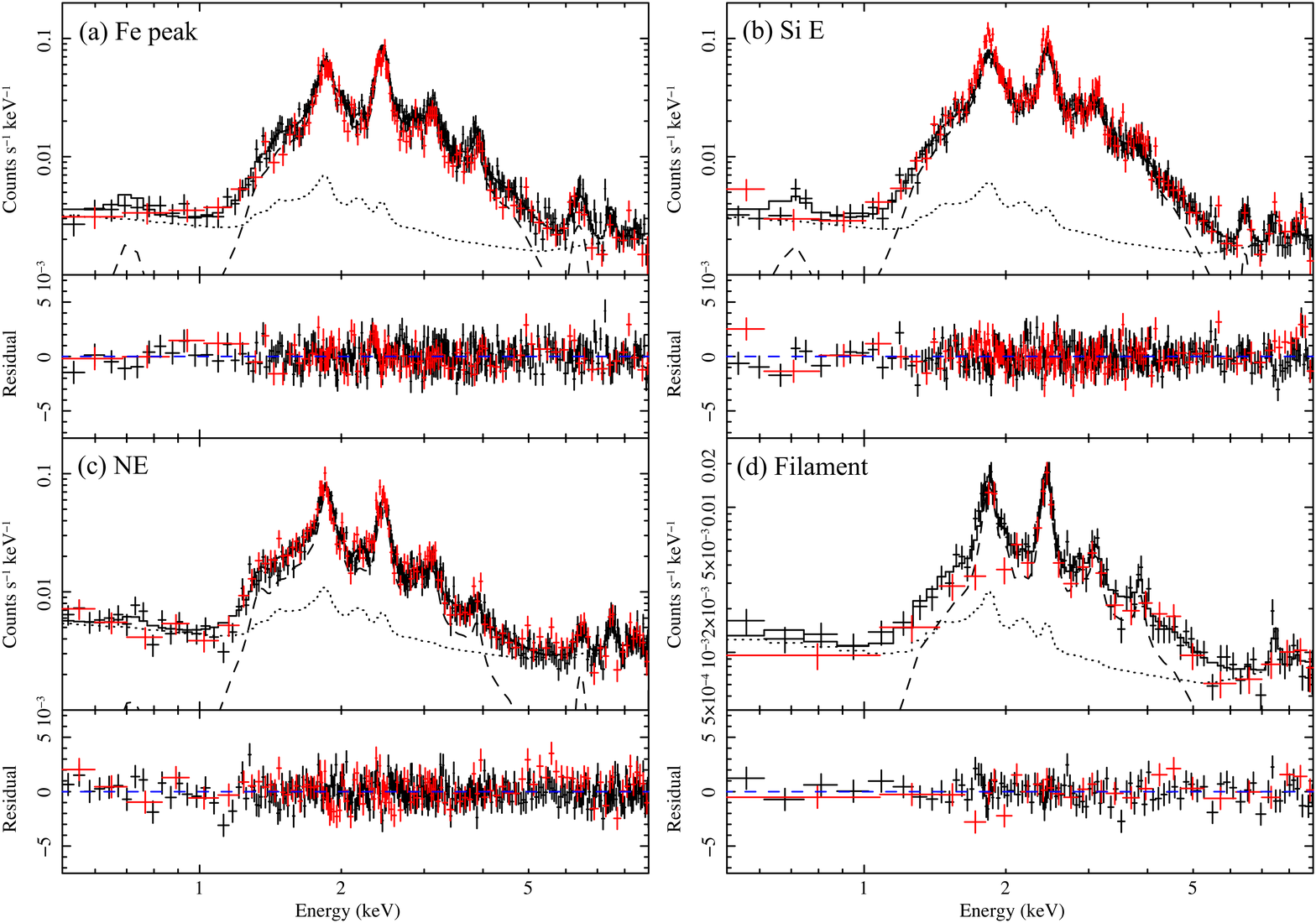}{0.75\textwidth}{}}
\caption{Spectra of representative regions: (a) Fe peak, (b) Si E, (c) NE, and (d) Filament defined in Figure\,\ref{fig:band}.
The black and red data are the data from the May and July observation series, respectively. The dashed lines 
represent the NEI component and Fe~K emission in the best-fit models for the spectrum taken in 2018 May, 
whereas the dotted lines represent the modeled background components. 
\label{fig:spec}}
\end{figure*}

\begin{deluxetable*}{cCCCCCC}[ht!]
\tablecaption{Best-fit Spectral Parameters for the Extended Emission \label{tab:parameter}}
\tablenum{3}
\tablewidth{0pt}
\tablehead{
\colhead{Components} &
\colhead{Parameters} &
\colhead{Fe} &
\colhead{Si N} &
\colhead{Si E} &
\colhead{Si SW} &
\colhead{N}
}
\startdata
Absorption\tablenotemark{a} & N_{\mathrm{H}}\ \mathrm{(10^{22}\ cm^{-2})} & 6.40^{+0.34}_{-0.24} & 6.68^{+0.29}_{-0.29} & 6.47^{+0.22}_{-0.20} & 6.52^{+0.18}_{-0.17} & 6.28^{+0.48}_{-0.50} \\
NEI plasma\tablenotemark{b} & kT_{\mathrm{e}}\ (\mathrm{keV}) & 1.10^{+0.04}_{-0.08} & 0.83^{+0.05}_{-0.04} & 0.93^{+0.03}_{-0.03} & 0.99^{+0.03}_{-0.03} & 0.81^{+0.08}_{-0.07} \\
 & \text{Mg} & 0.45^{+0.33}_{-0.24} & 1.14^{+0.40}_{-0.31} & 0.33^{+0.23}_{-0.19} & 0.63^{+0.21}_{-0.18} & 1.11^{+1.16}_{-0.60} \\
 & \text{Si} & 1.51^{+0.15}_{-0.16} & 1.42^{+0.14}_{-0.12} & 1.13^{+0.10}_{-0.94} & 1.48^{+0.10}_{-0.09} & 1.87^{+0.45}_{-0.17} \\
 & \text{S} & 2.28^{+0.15}_{-0.17} & 1.69^{+0.11}_{-0.11} & 1.49^{+0.09}_{-0.09} & 2.03^{+0.09}_{-0.08} & 2.13^{+0.27}_{-0.23} \\
 & \text{Ar} & 2.05^{+0.26}_{-0.24} & 2.18^{+0.29}_{-0.27} & 1.62^{+0.20}_{-0.19} & 2.14^{+0.16}_{-0.16} & 1.71^{+0.54}_{-0.49} \\
 & \text{Ca} & 4.26^{+0.58}_{-0.54} & 4.09^{+1.22}_{-0.96} & 1.85^{+0.44}_{-0.41} & 4.41^{+0.49}_{-0.47} & 1.82^{+2.05}_{-1.61} \\  
 & n_{\mathrm{e}}t\ \mathrm{(10^{11}\ s\ cm^{-3})} & 1.74^{+0.49}_{-0.36} & 1.48^{+0.37}_{-0.31} & 2.17^{+0.45}_{-0.38} & 1.49^{+0.21}_{-0.16} & 1.67^{+0.85}_{-0.63} \\ 
 & \text{VEM\tablenotemark{c}} & 8.22^{+1.87}_{-0.84} & 18.9^{+3.2}_{-2.9} & 17.5^{+1.8}_{-1.7} & 26.5^{+2.4}_{-2.4} & 5.90^{+1.96}_{-1.40} \\
Fe K & E\ (\mathrm{keV}) & 6.41^{+0.03}_{-0.03} & 6.48^{+0.06}_{-0.06} & 6.45^{+0.03}_{-0.03} & 6.40^{+0.03}_{-0.03} & 6.54^{+0.07}_{-0.07} \\
 & \sigma\ (\mathrm{eV}) & 206^{+35}_{-31} & 100\ (\text{fix}) & < 93 & 85^{+39}_{-40} & 100\ (\text{fix}) \\
 & \text{Flux}\tablenotemark{d} & 9.80^{+1.1}_{-1.1} & 1.65^{+0.65}_{-0.62} & 3.53^{+0.70}_{-0.66} & 6.77^{+1.16}_{-1.07} & 1.02^{+0.75}_{-0.68} \\
Photon Counts ($10^4$ conut) & & 1.75 & 1.83 & 2.05 & 4.22 & 1.24 \\
Background Contribution (\%) & & 18.4 & 19.1 & 16.4 & 21.4 & 48.8 \\
C-stat & & 2658.68 & 2690.62 & 2636.78 & 2739.07 & 2655.14 \\
dof & & 2311 & 2312 & 2311 & 2311 & 2312 \\ \hline\hline
 & & \text{NE} & \text{NW} & \text{E} & \text{SE} & \text{Filament} \\ \hline
Absorption\tablenotemark{a} & N_{\mathrm{H}}\ \mathrm{(10^{22}\ cm^{-2})} & 6.19^{+0.31}_{-0.29} & 5.30^{+0.75}_{-0.54} & 6.63^{+0.60}_{-0.43} & 6.69^{+0.57}_{-0.53} & 5.80^{+0.84}_{-0.47} \\
NEI plasma\tablenotemark{b} & kT_{\mathrm{e}}\ (\mathrm{keV}) & 0.79^{+0.07}_{-0.05} & 1.27^{+0.25}_{-0.25} & 0.88^{+0.06}_{-0.10} & 0.90^{+0.07}_{-0.07} & 1.41^{+0.24}_{-0.21} \\
 & \text{Mg} & 1.25^{+0.49}_{-0.36} & <1.70 & 1.18^{+0.81}_{-0.55} & 0.66^{+0.70}_{-0.47} & <1.23 \\
 & \text{Si} & 2.09^{+0.27}_{-0.22} & 3.03^{+0.89}_{-0.76} & 1.87^{+0.34}_{-0.28} & 1.61^{+0.31}_{-0.25} & 2.22^{+0.73}_{-0.47} \\
 & \text{S} & 2.00^{+0.17}_{-0.15} & 4.07^{+0.86}_{-0.80} & 2.49^{+0.27}_{-0.25} & 2.60^{+0.28}_{-0.24} & 2.67^{+0.53}_{-0.43} \\
 & \text{Ar} & 2.83^{+0.46}_{-0.40} & 3.72^{+1.03}_{-0.90} & 2.74^{+0.51}_{-0.47} & 3.16^{+0.48}_{-0.42} & 2.15^{+0.73}_{-0.64} \\
 & \text{Ca} & 5.73^{+2.04}_{-1.61} & 3.28^{+1.52}_{-1.30} & 5.22^{+1.07}_{-0.84}\ \tablenotemark{e} & 6.65^{+2.61}_{-1.58} & 4.39^{+1.82}_{-1.46} \\  
 & n_{\mathrm{e}}t\ \mathrm{(10^{11}\ s\ cm^{-3})} & 1.52^{+0.41}_{-0.35} & 1.59^{+1.21}_{-0.47} & 1.26^{+0.79}_{-0.33} & 1.46^{+0.57}_{-0.43} & 0.86^{+0.39}_{-0.23} \\ 
 & \text{VEM\tablenotemark{c}} & 11.1^{+2.3}_{-2.2} & 1.09^{+0.75}_{-0.35} & 7.04^{+3.05}_{-1.28} & 8.00^{+2.19}_{-1.76} & 0.91^{+0.41}_{-0.25} \\
Fe K & E\ (\mathrm{keV}) & 6.45^{+0.03}_{-0.03} & - & 6.46^{+0.03}_{-0.03}\ \tablenotemark{e} &
 6.46^{+0.03}_{-0.03}\ \tablenotemark{e}& - \\
 & \sigma\ (\mathrm{eV}) & 93.9^{+36}_{-35} & - & 106^{+35}_{-36} & 167^{+71}_{-58} & - \\
 & \text{Flux}\tablenotemark{d} & 5.22^{+0.94}_{-0.86} & - & 5.31^{+1.10}_{-1.01} & 3.88^{+1.10}_{-1.00} & - \\
Photon Counts ($10^4$ conut) & & 3.73 & 0.69 & 1.48 & 1.52 & 0.40 \\
Background Contribution (\%) & & 35.5 & 45.6 & 46.5 & 40.6 & 35.8 \\
C-stat & & 2704.56 & 2694.10 & 2601.55 & 2643.46 & 2671.64 \\
dof & & 2311 & 2314 & 2313 & 2312 & 2314 \\
\enddata
\tablenotetext{a}{The absorption cross sections are taken from \cite{Verner96}. 
}
\tablenotetext{b}{Solar abundances of \cite{Wilms00} 
are assumed.}
\tablenotetext{c}{In units of $10^{11}\ \mathrm{cm^{-5}}$.
The volume emission measure (VEM) is given as $\int n_{\mathrm{e}}n_{\mathrm{H}}dV/(4\pi D^2)$,
where $V$ and $D$ are the volume of the emission region (cm$^3$)
and the distance to the emitting source (cm), respectively.}
\tablenotetext{d}{In units of $10^{-6}\ \mathrm{photon\ cm^{-2}\ keV^{-1}\ s^{-1}}$.}
\tablenotetext{e}{These are determined using spectra extracted from a larger region containing both E and SE, 
since the values cannot be constrained for each individual region.}
\end{deluxetable*}

Here we analyze ACIS-I spectra extracted from the regions given in Figure\,\ref{fig:band}, 
where the point sources detected in the previous section (Table\,\ref{tab:srclist}) are excluded.
In our analysis we model the background emission instead of subtracting it.
Figure\,\ref{fig:spec} shows the spectra of several representative regions, 
where black and red are the data from the May and July observations, respectively. 
The emission lines of the IMEs and Fe are clearly resolved in each region. 
Background data are extracted from an annulus region surrounding the SNR with 
inner and outer radii of $5'.0$ and $7'.0$, respectively (Figure\,\ref{fig:snr}). 
We model and fit the background spectra simultaneously with the source spectra, 
instead of subtracting them from the source data,
to properly take into account the difference in the effective area
between the source and background regions.
The details of the background spectral modeling are presented in Appendix \ref{sec:background}. 
In the following spectral analysis, we fit unbinned data using the C-statistics \citep{Cash79} 
to estimate the model parameters and their error ranges without bias, 
although the binned spectra are shown in the figures for clarity.

We first fit the spectra with a model of an optically thin thermal plasma in the non-equilibrium 
ionization (NEI) state, i.e., the {\tt vnei} model in the XSPEC package, based on the latest 
atomic database ({\tt AtomDB}\footnote{\url{http://www.atomdb.org}}  version 3.0.9).
We use the {\tt tbnew\_gas} model
\footnote{\url{https://pulsar.sternwarte.uni-erlangen.de/wilms/research/tbabs/}} 
for the foreground absorption, 
assuming the photoelectric absorption cross sections taken from \cite{Verner96}. 
The free parameters are the absorption column density $N_{\mathrm{H}}$,
electron temperature $kT_e$, ionization timescale $n_et$,
volume emission measure (VEM), and elemental abundances of Mg, Si, S, Ar, Ca,
and Fe relative to the solar abundance table of \cite{Wilms00}. 
This model gives good fits to the spectra below 5 $\mathrm{keV}$, but fails to reproduce 
the Fe~K emission. We find that the model generally predicts an Fe~K centroid energy 
higher than the observed values, which indicates that the Fe~K emission originates from 
another plasma component with a lower ionization state.

We thus add a Gaussian component to reproduce the K-shell emission from the low-ionized Fe. 
The addition of this component significantly improves the fit, with the C-stat value reduced
from 1003.00/805 to 920.43/803 in the 5.0--8.0\,keV band, including the strong Fe~K line.
The best-fit model spectra and parameters are given 
in Figure\,\ref{fig:spec} and Table\,\ref{tab:parameter}, respectively. 
The abundances of the IMEs (but for Mg) are generally higher than the solar values, 
implying a significant ejecta contribution to the {\tt vnei} component. 
We also find that both the electron temperature (of the {\tt vnei} component) and 
the Fe K centroid energy vary among the regions. 
The lowest temperature is observed in the N and NE, consistent with the mean photon 
energy map (Figure\,\ref{fig:band}b). The Fe K centroid energy is marginally lower
in the Fe peak region than in most of the outer regions,
indicating that the Fe ejecta in the former has a low ionization degree. 
The absorption column density is found to be almost uniform throughout the SNR, 
$\sim 6.5\times 10^{22}\ \mathrm{cm^{-2}}$. This value is slightly higher than the previous 
measurements of $N_{\mathrm{H}}=$ 4.0--5.5 $\times 10^{22}\ \mathrm{cm^{-2}}$ \citep{Yamauchi05,Combi10,Giacani11,Yamaguchi12b}. 
This discrepancy is likely due to the difference in the reference solar abundance; 
the previous work referred to \cite{AG89}, while we refer to \cite{Wilms00}. 
In fact, if we fit the ACIS-I spectra using the solar abundance table of \cite{AG89}, 
we obtain $N_{\mathrm{H}}\sim 4.5\times 10^{22}\ \mathrm{cm^{-2}}$, 
consistent with the previous work.


\begin{deluxetable}{CCCC}[ht!]
\tablecaption{Best-fit Spectral Parameters for the pure Fe Component \label{tab:parameter2}}
\tablenum{4}
\tablewidth{0pt}
\tablehead{
\colhead{Parameters} &
\colhead{Fe Peak} &
\colhead{Si-arc} &
\colhead{Rim}
}
\startdata
kT_e\ ({\rm keV}) & 2.9^{+0.1}_{-0.1} & 5.6^{+5.6}_{-0.4} & 3.8^{+0.2}_{-0.2} \\
 n_{{\rm e}}t\ \mathrm{(10^{9}\ s\ cm^{-3})} & <0.14 & 3.5^{+1.8}_{-1.8} & 17.4^{+2.1}_{-2.2} \\
\text{VEM\tablenotemark{a}}\ (10^{5}\ {\rm cm^{-5}})& 1.2^{+0.1}_{-0.1} & 0.8^{+0.4}_{-0.1} & 2.3^{+0.1}_{-0.5} \\
\enddata
\tablenotetext{a}{The VEM is given as $\int n_{\rm e}n_{\rm Fe}dV/(4\pi D^2)$.}
\end{deluxetable}

Next, we replace the Gaussian component with another {\tt vnei} component
to account for the emission from pure Fe ejecta.
The SNR is now divided into three regions, the Fe peak, Si-arc (containing Si E, N, and SW),
and rim (containing N, NE, NW, SE, S, and Filament),
and their spectra are jointly fitted with a common $N_{\rm H}$ value.
Table\,\ref{tab:parameter2} gives the parameter values obtained for the pure Fe ejecta component.
We find that the temperature of this component ($kT_e>2.8$\,keV) is significantly
hotter than that of the IME component ($kT_e=$ 0.8--1.1\,keV).
We also find that $n_e t$ of the Fe ejecta $<2 \times10^{10}$\,s\,cm$^{-3}$ is lower than
that of the IME component $\sim2 \times10^{11}$\,s\,cm$^{-3}$,
and that the lowest $n_e t$ of the Fe ejecta component is achieved at the Fe peak region.
\begin{figure*}[ht!]
\gridline{\fig{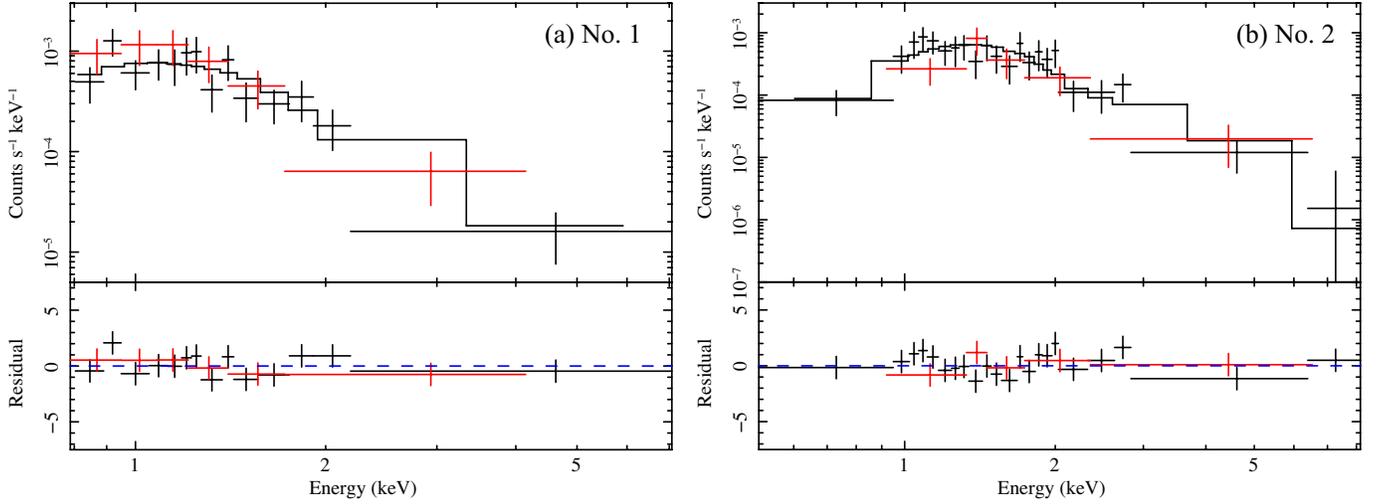}{1.0\textwidth}{}}
\caption{Background-subtracted spectra of the source of (a) No.\,1 and (b) No.\,2. \label{fig:cxou_spec}
The black and red data are the data from the May and July observation series, respectively. 
The solid line represents the best-fit power-law model for the former spectrum.}
\end{figure*}


\subsubsection{Point-like Sources}

We perform spectral analysis of two bright sources (No.\,1 and No.\,2) detected in the soft 
X-ray band, both embedded in the diffuse emission from the SNR (Figure\,\ref{fig:band}e).
The spectra of source No.\,1 (CXOU~J170357.8--414302) and No.\,2 are shown 
in Figure\,\ref{fig:cxou_spec}. We fit the spectra of both sources with the following three models: 
an absorbed power law ({\tt tbnew\_gas} $\times$ {\tt powerlaw}),
an absorbed blackbody ({\tt tbnew\_gas} $\times$ {\tt bbody}), and an absorbed 
optically thin thermal plasma with the solar abundances ({\tt tbnew\_gas} $\times$ {\tt apec}).
We use background spectra extracted from the annulus surrounding each source region.
For source No.\,1, we obtain 160 counts for the background-corrected spectrum
with 16\% background contribution, and 136 counts with 8.6\% for No.\,2.
The best-fit parameters we obtain are given in Table\,\ref{tab:two_src_par}.
The estimated column densities are significantly lower than that of the SNR
($N_{\mathrm{H}}$ $\approx$ $6.5\times 10^{22}\ \mathrm{cm^{-2}}$) for any model assumption,
consistent with the foreground stellar object explanation as suggested by \cite{Combi10} for No.\,1.
The spectra of the other detected sources (Table\,\ref{tab:srclist}) with sufficient photon counts 
are analyzed as well, obtaining statistically acceptable results (C-stat/dof $\lesssim$ 1.2) 
for all the sources. The results are discussed in more detail in \S\ref{subsec:cco}.

\begin{deluxetable}{cCCCCCC}[ht!]
\tablecaption{Best-fit spectral parameters for the bright point sources\label{tab:two_src_par}}
\tablenum{5}
\tablewidth{0pt}
\tablehead{
\colhead{No.} &
\colhead{$N_{\mathrm{H}}$\tablenotemark{a}} &
\colhead{$\Gamma$} &
\colhead{$N_{\mathrm{H}}$\tablenotemark{a}} &
\colhead{$kT_{\mathrm{bb}}$\tablenotemark{b}} &
\colhead{$N_{\mathrm{H}}$\tablenotemark{a}} &
\colhead{$kT_{\mathrm{APEC}}$\tablenotemark{b}}
}
\startdata
1 & <0.08 & 3.44^{+0.29}_{-0.24} & <0.06 & 0.24^{+0.02}_{-0.01} & 0.58^{+0.18}_{-0.18} & 0.92^{+0.10}_{-0.14} \\
2 & 0.93^{+0.35}_{-0.31} & 3.84^{+0.53}_{-0.47} & <0.16 & 0.43^{+0.03}_{-0.03} & 1.29^{+0.25}_{-0.50} & 1.04^{+0.31}_{-0.15} \\
\enddata
\tablenotetext{a}{In units of $10^{22}\ \mathrm{cm^{-2}}$.}
\tablenotetext{b}{In units of keV.}
\end{deluxetable}



\section{Discussion}
\label{sec:discuss}

\subsection{Stratified Elemental Composition}
\label{subsec:discussion_composition}


Thanks to the deep observations with Chandra, we have revealed the centrally peaked 
Fe K emission surrounded by the arc-like structure of the IME emission. 
This radially stratified chemical composition with Fe at the interior is consistent with 
the standard picture of SN Ia nucleosynthesis dominated by the carbon detonation
\citep[e.g.][]{Iwamoto99,Seitenzahl13}. 
Several other SN Ia explosion models, such as those involving pure carbon deflagration 
\citep[e.g.][]{Fink14} or He-shell detonation \citep[e.g.][]{Sim12}, predict  
a substantial amount of $^{56}$Ni (which decays into Fe) synthesized outside of the IME.
We do not find such evidence in the X-ray data of G344.7--0.1,
which may rule out these explosion scenarios as the origin of this SNR.
We note that our spectral data below 1\,keV, where Fe~L lines fall, are missing due to the strong absorption,
leaving the possibility that a significant amount of Fe is mixing into the IME-rich layer.
Unless this is the case, our results prefer the carbon detonation model for the origin of G344.7--0.1.

\begin{figure}[h!]
\gridline{\fig{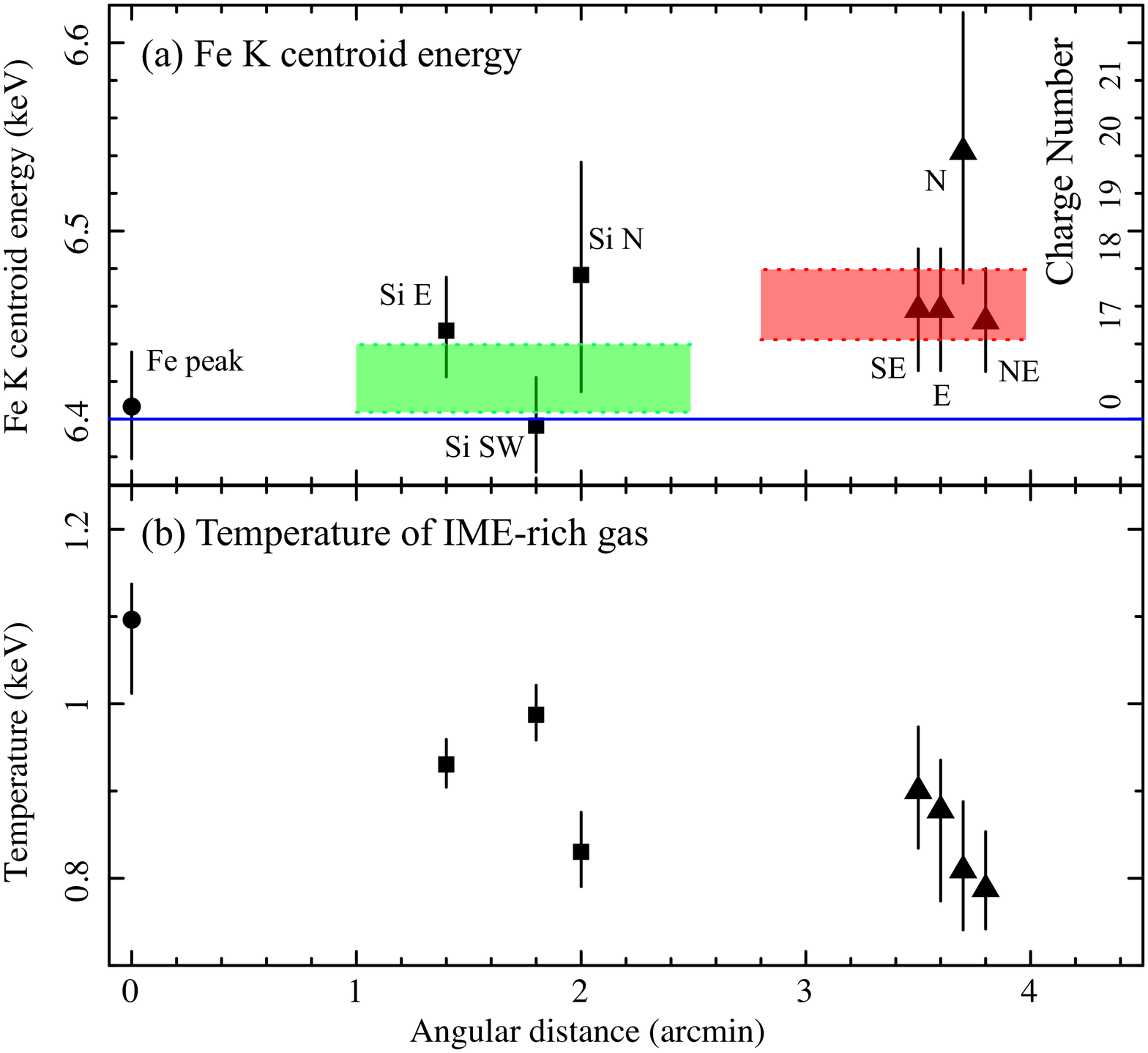}{1.0\columnwidth}{}}
\caption{(a) Fe~K centroid energy as a function of the radial distance from the Fe peak location.
The corresponding charge number of Fe ions is given on the right.
Green and red areas represent the values constrained by analyzing spectra from larger regions 
containing the entire Si arc (where square plots of Si E, SW, and N are all merged) and the entire rim regions 
(where triangle plots of SE, E, N, and NE are all merged), respectively. 
The blue solid line indicates the 6.40\,keV level that corresponds to neutral Fe.
(b) Electron temperature for the IME-rich gas component as a function of the radial distance from the Fe peak location.
Square plots are the merged Si-arc region, and the triangle ones are the rim region.
 \label{fig:fecent}}
\end{figure}

We have also found that the peak location of the Fe emission (and thus the centroid of the IME arc) 
is offset to the west by $\sim1'.6$ (corresponding to $\sim 2.8$\,pc at the distance of 6\,kpc)
with respect to the geometric center of the SNR. This implies that either 
the SN explosion was asymmetric, or the Fe K peak location is the actual explosion center 
but an apparent asymmetry has formed during the SNR evolution. 
The spatial distribution of the plasma conditions may help disentangle this degeneracy. 
In particular, the ionization state of the Fe ejecta that is inferred from the Fe K centroid energy 
offers a useful diagnostic of the time duration since the ejecta were heated by the reverse shock 
\citep[e.g.][]{Badenes06,Yamaguchi14a}.
Figure\,\ref{fig:fecent}(a) shows the Fe~K centroid energy observed in each region
as a function of the angular distance from the location of the Fe emission,
which reveals a positive correlation between the two quantities.
The mean values of the centroid energy in the Fe peak and Si arc regions
correspond to the charge states of Fe$^{10+}$ and Fe$^{17+}$, respectively.
Similarly, Table\,\ref{tab:parameter2} indicates that a lower ionization timescale
is achieved in the Fe peak region than in the outer regions.
This suggests that the Fe ejecta at the inner regions were heated by the reverse shock more recently.
We also plot in Figure\,\ref{fig:fecent}(b) the electron temperatures of the NEI component 
of each region. In contrast to the Fe K centroid, the temperature is anticorrelated with 
the distance from the Fe peak location.
We thus conclude that the location of the Fe peak region is the true explosion center of this SNR. 
The origin of the asymmetric morphology is discussed in more detail in \S\ref{subsec:discussion_asymmetry}. 


One caveat is that the variation in the Fe K centroids may also be contributed by 
the line-of-sight velocity of the Fe ejecta, as was claimed for the Kepler SNR 
\citep{Kasuga18}. 
We find that the observed variation in the Fe K centroid ($\sim$50\,eV) requires
a line-of-sight velocity $v_{\rm sight}$ of 2400\,km\,s$^{-1}$ for the Fe ejecta at the Fe peak region. 
If the Fe ejecta of the Fe peak location have been freely expanding for 3000\,yr
with $v_{\rm sight} \sim$2400\,km\,s$^{-1}$, the ejecta reach 7.4\,pc
from the SN explosion center, larger than the radius of this SNR (5\,pc; assuming a distance of 6\,kpc).
Given that the ejecta must have been decelerated by the reverse shock,
we should take this estimate as a lower limit.
Since an unnaturally large asymmetry in the explosion is required
to explain this line shift solely with the Doppler effect,
the ionization effect is likely to be dominant in
the observed variation in the Fe~K centroid energies.
The degeneracy between the ionization effect and Doppler shift in the Fe K emission can be 
disentangled by measurement of the Fe K$\beta$/K$\alpha$ flux ratio that strongly depends on 
the dominant charge state and thus constrains the rest-frame Fe K centroid \citep{Yamaguchi14b}. 
Unfortunately, neither our Chandra observation nor any previous work allows us to 
detect Fe K$\beta$ emission due to poor photon statistics and limited energy resolution.
Future high-resolution spectroscopy with XRISM and Athena are crucial to determine both 
charge state and line-of-sight velocity accurately. 



\begin{figure*}[ht!]
\gridline{\fig{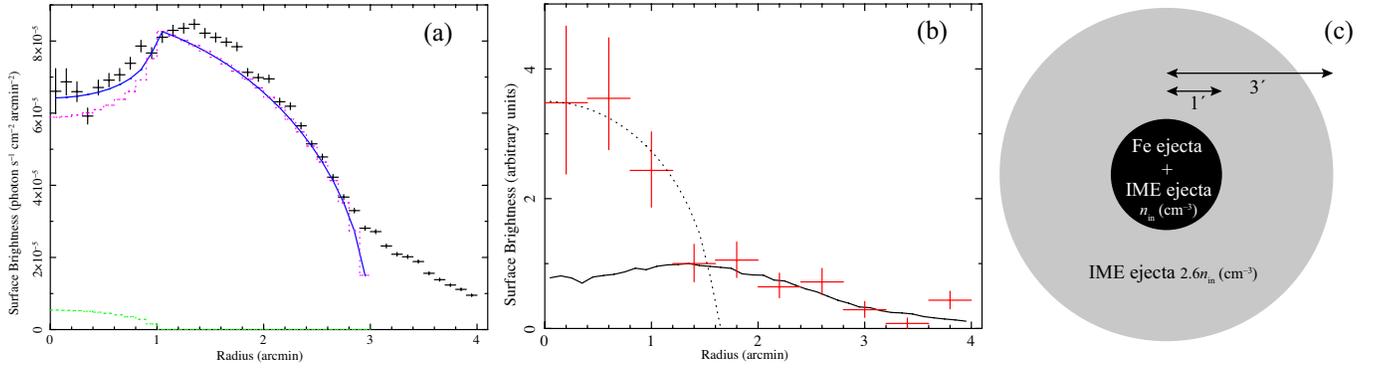}{1.0\textwidth}{}
}
\caption{(a) Radial profile of the surface brightness in the IME band ($1.5$--$5.0\ \mathrm{keV}$), 
where the Fe peak position is assumed to be the SNR center. 
The blue solid line shows the modeled surface brightness that assumes a low-density inner sphere with 
a radius of $1'$ and a thick high-density shell with inner and outer radii of $1'$ and $3'$, respectively
(see the text for more details).
The density ratios of the outer shell to the inner sphere are assumed to be $n_{\rm shell}/ n_{\rm in}$ = 2.6. 
The magenta and green dashed lines show the contribution of the outer shell and inner sphere components 
to the total surface brightness, respectively.
(b) Linearly scaled radial profiles of (a) and of surface brightness in the Fe~K emission band.
The solid line is for the scaled surface brightness of (a), and the red crosses for Fe~K emission.
The dotted line represents the modeled surface brightness assuming a spherical distribution of Fe ejecta.
The continuum-subtracted
flux of Fe~K emission is determined by fitting with the model in \S\ref{subsec:ext}.
(c) Schematic view of the density distribution of G344.7--0.1
indicated in (a) and (b). \label{fig:radmodel}} 
\end{figure*}

Now we estimate the three-dimensional distribution of the {\tt vnei} component that is dominated 
by the IME ejecta, assuming that the Fe peak region is the true explosion center. Since the surface-brightness
profiles of Si, S, Ar, and Ca (Figure\,\ref{fig:surbri}) are virtually identical with one another, 
we use a radial profile of the 1.5--5.0\,keV emission as representative of the IME distribution. 
Figure\,\ref{fig:radmodel}(a) shows the extracted radial profile compared with a simple two-zone 
spherical model assuming a uniform density and uniform abundance distribution in each zone.
We also take into account the temperature gradient, and thus the emissivity variation, 
between the two zones, where $\sim$\,1.1\,keV at $<1'$ and $\sim$\,0.9\,keV at $1'$--$3'$.
Figure\,\ref{fig:radmodel}(b) shows the continuum-subtracted surface brightness of the Fe~K
and 1.5--5.0-keV emission, suggesting an enhanced density of Fe ejecta at $<1'$.
We find that the observed profile is well reproduced by a model
schematized in Figure\,\ref{fig:radmodel}(c): 
a low-density ($n_{\rm in}$) inner sphere with a radius of $1'$ and 
a thick high-density ($n_{\rm shell}$ = $2.6 n_{\rm in}$) shell 
with inner and outer radii of $1'$ and $3'$, respectively. The IME arc structure found in Figure\,\ref{fig:band}(c) 
corresponds to the inner part of the thick shell, where the highest surface brightness is achieved. 
Note that the angular distance between the explosion center and the west rim is $\sim 3'$, 
and thus the excess in the brightness beyond this radius (Figure\,\ref{fig:radmodel}a) 
is solely due to the emission extended to the northeast.


\subsection{Asymmetry}
\label{subsec:discussion_asymmetry}


As discussed in the previous section, the identified explosion center is offset from the geometric 
center of the SNR. Therefore, the X-ray morphology of G344.7--0.1 is relatively asymmetric.
\cite{Lopez11} classified this SNR as a possible CC SNR because of such highly asymmetric X-ray morphology,
based on the analysis of the power-ratio method (PRM).
However, the dataset used in \cite{Lopez11} has more limited photon statistics
than our 200 ks Chandra data.
Thus, we apply the PRM on our new dataset of G344.7--0.1 to measure the asymmetry of this SNR.
We replicate the analysis from \cite{Lopez11}, where the PRs are calculated
on the 0.5--2.1\,keV band exposure-corrected image of SNR, and obtain $P_2/P_0=9.84^{+2.05}_{-2.32}$,
$P_3/P_0=1.75^{+0.55}_{-0.62}$\footnote{High $P_2/P_0$ indicates elliptical or elongated morphology,
and high $P_3/P_0$ indicates asymmetric or nonuniform surface-brightness distribution \citep{Lopez11}.}.
The values are now comparable with those for the other well-established Type Ia SNRs, such as Kepler.
We note that the surface-brightness-weighted geometric center,
i.e. the aperture center for the PRM, is
($\alpha_{\rm J2000}$, $\delta_{\rm J2000}$)=(17$^{\rm h}$03$^{\rm m}$55$^{\rm s}$.5, --41$^{\circ}$42$'$54$''$.5),
which is almost the same as the geometric center of the SNR, and thus
is offset from our proposed explosion center of
($\alpha_{\rm J2000}$, $\delta_{\rm J2000}$)=(17$^{\rm h}$03$^{\rm m}$49$^{\rm s}$.1, --41$^{\circ}$42$'$46$''$.0).
This offset and apparent asymmetry seems to be caused by the nonuniform
distribution of the ambient medium. In fact, other SNRs expanding in such ambient medium distributions
(e.g., Kepler, N103B) also show high asymmetric morphology and off-center aperture center
like G344.7--0.1 \citep{Lopez11}.

\begin{figure}[h!]
\gridline{\fig{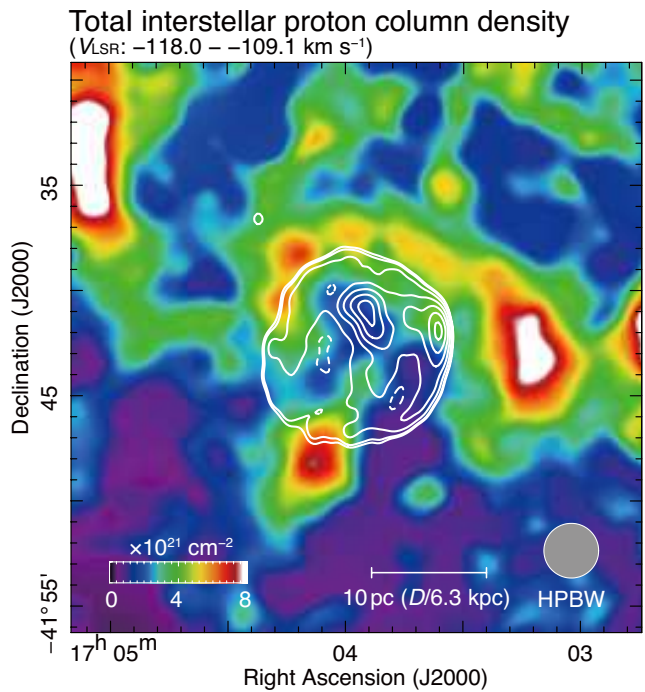}{1.0\columnwidth}{}}
\caption{Map of total proton column density toward G344.7--0.1.
The integration velocity range is from --118.0\,km\,s$^{-1}$ to --109.1\,km\,s$^{-1}$,
corresponding to the distance of $\sim$\,6.3\,kpc. The half-power beamwidth is shown by field circle.
The scale bar is calculated by assuming the distance of 6.3\,kpc.
The white contours indicate a 843\,MHz radio continuum obtained with MOST \citep{WG96}.
The contour levels are 10.0, 12.5, 20.0, 32.5, 50.0, 72.5, and 100.0\,mJy\,beam$^{-1}$.
\label{fig:co-h1_map}} 
\end{figure}

The nonuniform density structure of the ambient medium is indeed found in our radio data.
Figure\,\ref{fig:co-h1_map} shows the total interstellar proton column density map $N_{\rm p}{\rm (H_2+HI)}$
combining both the molecular (CO: NANTEN2) and atomic \citep[HI:][]{McClure05} components.
The details of the radio observations and data analysis are described in Appendix \ref{sec:radio}.
The velocity range of the gas density map is from --118.0\,km\,s$^{-1}$ to --109.0\,km\,s$^{-1}$,
which corresponds to the kinematic distance of 6.2--6.4\,kpc
adopting a model of the Galactic rotation curve by \cite{BB93}.
The velocity range is determined by CO/HI cavities in the position-velocity diagrams
(see Appendix \ref{subsec:expansion}), suggesting an expanding gas motion originated
by accretion winds from the progenitor system of the SNR \citep[e.g.,][]{Sano17,Sano18}.
Therefore, the molecular and atomic clouds at the velocity range are likely associated with the SNR.
The velocity range and distance to the SNR are consistent with the previous HI study by \cite{Giacani11}.

We find denser clouds possibly interacting with the SNR in the west (Figure\,\ref{fig:co-h1_map}), 
consistent with the previous results from the Mopra Southern Galactic Plane CO Survey
\citep{Burton13,Braiding15,Lau19}. 
The SNR expansion is likely to be decelerated by this interaction toward the west, resulting in 
the asymmetric morphology we observe. This interpretation is supported by the temperature 
gradient inferred from the mean energy map (Figure\,\ref{fig:band}b); the lower temperature is achieved in 
the northeast region probably due to more efficient adiabatic expansion. A similar temperature 
gradient due to the asymmetric ambient density distribution is observed in other middle-aged SNRs, 
e.g., W49B \citep{Lopez13,Yamaguchi18}. Alternatively, the SN ejecta at the west could have 
been reheated by reflection shocks originating from the interaction with the dense clouds, as was 
observed in, for instance, Kes\,27 \citep{Chen08}. 

We have also discovered a local asymmetric structure at the west rim, ``Filament'' in Figure\,\ref{fig:band}(c).
Interestingly, the measured IME abundances are enhanced with respect to the solar values.
Moreover, the relative abundances of S, Ar, Ca to Si are comparable to the solar composition
and that of Mg is substantially lower, which is consistent with the typical nucleosynthesis yields
of the incomplete Si burning \citep[e.g.,][]{Iwamoto99,Seitenzahl13,Townsley16}.
This implies an ejecta origin of this structure,
despite that its filamentary morphology and location being more consistent with
an interpretation of the forward shock front that propagates into the ISM.
If the structure is indeed dominated by the SN ejecta, it is reminiscent of the ejecta knots in {\it Tycho} SNR, 
suggesting that such local asymmetry is somewhat common in a certain type of Type Ia SNRs. 

\subsection{Point-Like Sources}
\label{subsec:cco}

\begin{deluxetable*}{cCCCC}[h!]
\tablecaption{Spectral Properties of All Point Sources \label{tab:pointsrc_fit}}
\tablenum{6}
\tablewidth{0pt}
\tablehead{
\colhead{No.\tablenotemark{a}} &
\colhead{$N_{\mathrm{H}}\ \mathrm{(10^{22}\ cm^{-2})}$} &
\colhead{$kT_{\mathrm{bb}}\ \mathrm{(keV)}$} &
\colhead{Luminosity\tablenotemark{b} ($10^{33}\ \mathrm{erg\,s^{-1}}$)} &
\colhead{Distance $\mathrm{(arcmin)}$\tablenotemark{c}}
}
\startdata
1 & <0.06 & 0.24^{+0.02}_{-0.01} & 4.85^{+0.93}_{-0.47}\times 10^{-2} & 1.66  \\
2 & <0.16 & 0.43^{+0.03}_{-0.03} & 3.53^{+0.56}_{-0.32}\times 10^{-2} & 3.34  \\
3 & <0.09 & 0.58^{+0.06}_{-0.05} & 2.58^{+0.39}_{-0.35}\times 10^{-2} & 3.99 \\
6 & <0.10 & 0.59^{+0.07}_{-0.07} & 1.02^{+0.21}_{-0.16}\times 10^{-2} & 4.69  \\
7 & <0.38 & 0.28^{+0.10}_{-0.07} & 1.46^{+1.81}_{-0.30}\times 10^{-2} & 4.45  \\
8 & <0.10 & 0.84^{+0.14}_{-0.15} & 1.11^{+0.37}_{-0.29}\times 10^{-2} & 5.35  \\
9 & <0.93 & 0.16^{+0.04}_{-0.06} & 1.86^{+12.3}_{-0.87}\times 10^{-2} & 2.93  \\
11 & 1596^{+616}_{-1008} & <0.24\tablenotemark{d} & <3.15\times 10^{16} & 5.11  \\
14 & <0.38 & 0.34^{+0.07}_{-0.08} & 0.84^{+0.62}_{-0.17}\times 10^{-2} & 6.55  \\
15 & <0.14 & 0.48^{+0.05}_{-0.05} & 3.23^{+0.61}_{-0.48}\times 10^{-2} & 6.33 \\
16 & <0.68 & 0.21^{+0.05}_{-0.04} & 1.05^{+0.80}_{-0.38}\times 10^{-2} & 4.29  \\
17 & <0.71 & 0.54^{+0.08}_{-0.07} & 4.08^{+1.01}_{-0.23}\times 10^{-2} & 2.66  \\
18 & <2.45 & 0.25^{+0.10}_{-0.14} & 0.78^{+10.6}_{-0.37}\times 10^{-2} & 3.85  \\
19 & <0.62 & 0.45^{+0.17}_{-0.34} & 0.23^{+0.24}_{-0.23}\times 10^{-2} & 3.09  \\
20 & <0.25 & 0.80^{+0.07}_{-0.08} & 4.66^{+0.69}_{-0.60}\times 10^{-2} & 1.04  \\
22 & <0.54 & 0.39^{+0.10}_{-0.10} & 1.05^{+0.81}_{-0.20}\times 10^{-2} & 4.38  \\
24 & <1.23 & 0.56^{+0.11}_{-0.16} & 1.49^{+1.27}_{-0.36}\times 10^{-2} & 4.17 \\
25 & <0.41 & 0.88^{+0.44}_{-0.31} & 0.74^{+0.44}_{-0.28}\times 10^{-2} & 4.82  \\
26 & 2.65^{+1.54}_{-1.20} & 0.79^{+0.17}_{-0.13} & 4.05^{+1.52}_{-0.90}\times 10^{-2} & 3.03  \\
27 & 38^{+0.8}_{-4.7} & 2.73^{+0.14}_{-0.02}\times 10^{-2} & <3.60\times 10^{29} & 2.63  \\
28 & <0.62 & 0.48^{+0.15}_{-0.14} & 0.51^{+0.31}_{-0.17}\times 10^{-2} & 4.58  \\
29 & <1.90 & >2.26 & <253 & 6.82  \\
30 & <0.39 & 0.90^{+0.27}_{-0.19} & 0.66^{+0.38}_{-0.35}\times 10^{-2} & 3.96  \\
31 & 7.44^{+5.60}_{-3.29} & 0.37^{+0.13}_{-0.11} & <2.77 & 1.51  \\
& 6.69\ \tablenotemark{e} &0.39^{+0.05}_{-0.04}&0.21^{+0.09}_{-0.08}&\\
32 & 8.66^{+7.79}_{-3.38} & <0.18 & <7.64\times 10^{10} & 1.74  \\
33 & 38^{+23}_{-14} & 0.07^{+0.04}_{-0.02} & <9.00\times 10^{11} & 3.17 \\
34 & 2.42^{+1.19}_{-0.89} & 0.62^{+0.09}_{-0.08} & 0.12^{+0.05}_{-0.03} & 2.89  \\
35 & >710 & <197 & <5.36\times 10^{7} & 1.83  \\
36 & 6.74^{+3.17}_{-2.49} & 0.35^{+0.09}_{-0.07} & <1.12 & 2.93  \\
& 6.63\ \tablenotemark{e} &0.35^{+0.06}_{-0.03}&0.24^{+0.10}_{-0.13}&\\
37 & 2.18^{+1.90}_{-1.29} & 0.63^{+0.20}_{-0.14} & 3.27^{+2.96}_{-1.18}\times 10^{-2} & 1.52 \\
38 & 27^{+19}_{-13} & 0.12^{+0.07}_{-0.04} & <4.12\times 10^{5} & 1.45 \\
39 & 3.60^{+2.73}_{-1.78} & 0.49^{+0.14}_{-0.12} & 5.67^{+12.1}_{-2.97}\times 10^{-2} & 2.80 \\
41 & <28.3 & <4.19 & 1.01^{+8.50}_{-0.55}\times 10^{-2} & 1.33 \\
42 & 3.27^{+6.25}_{-3.11} & 0.67^{+0.44}_{-0.32} & 1.06^{+8.65}_{-0.64}\times 10^{-2} & 0.97  \\
43 & 4.31^{+2.90}_{-2.00} & 0.34^{+0.13}_{-0.10} & <1.47 & 2.29 \\
& 4.76\ \tablenotemark{e} &0.33^{+0.04}_{-0.09}&0.25^{+0.72}_{-0.07}&\\
45 & 77^{+21}_{-17} & 3.87^{+4.57}_{-3.57}\times 10^{-2} & <3.60\times 10^{29} & 1.97  \\
46 & 122^{+16}_{-34} & <0.06 & <416 & 2.12 \\
47 & 2.25^{+1.80}_{-1.41} & 0.50^{+0.18}_{-0.15} & 2.93^{+5.54}_{-1.58}\times 10^{-2} & 2.50 \\
48 & 7.80^{+6.99}_{-3.60} & 0.29^{+0.14}_{-0.13} & <5.13 & 4.94 \\
& 6.50\ \tablenotemark{e} &0.33^{+0.05}_{-0.04}&0.19^{+0.12}_{-0.10}&\\
49 & <11 & 0.52^{+0.99}_{-0.18} & <0.37 & 4.22 \\
\enddata
\tablenotetext{a}{Faint sources are excluded (see \S\ref{subsec:img}).}
\tablenotetext{b}{We assume the distance to the SNR to be 6\,kpc.}
\tablenotetext{c}{The angular distance from the Fe peak location.}
\tablenotetext{d}{We allow $kT_{\rm bb}$ to vary in 0.2--0.5\,keV for this source.}
\tablenotetext{e}{Best-fit values applied constraints in error ranges on each corresponding SNR region in Table\,\ref{tab:parameter}.}
\end{deluxetable*}


We have analyzed the spectra of the point-like sources detected with the total photon counts of 30 or 
more (see Table\,\ref{tab:srclist}). The best-fit parameters for the absorbed blackbody model are given 
in Table\,\ref{tab:pointsrc_fit}, where the angular distance from the explosion center to the sources is 
given as well. If the sources are physically associated with the SNR, the measured absorption column 
density must be consistent with that of the SNR (5--7 $\times 10^{22}$\,cm$^{-2}$). We find that only 
eight sources, No.\,31, 36, 39, 41, 42, 43, 48, and 49, satisfy this criterion, and the other sources are 
most likely either foreground or background objects.

Next, we compare the temperature and luminosity of these sources with typical values for CCOs: 
$kT_{\rm bb}$ = 0.2--0.5\,keV and $10^{33-34}$\,erg\,s$^{-1}$ \citep[e.g.][]{Pavlov04,Gotthelf13}.
We find that four sources, No.\,31, 36, 43 and 48 show typical $kT_{\rm bb}$ and luminosity as CCOs.
These sources, however, have relatively large errors of 37--90\% in $N_{\mathrm{H}}$.
Thus, we allow their $N_{\mathrm{H}}$ values to vary in the error ranges of $N_{\mathrm{H}}$
for the corresponding regions in Table\,\ref{tab:parameter} for each source.
We finally find that all these four sources have low luminosities with expected $N_{\mathrm{H}}$ values
of the different SNR regions. Therefore we conclude that there is no CCO associated with G344.7--0.1.

\section{Conclusions}
\label{sec:con}

The elemental distribution in SNRs is the key to understanding the explosion mechanism of SNe. 
We have presented the results of the deep Chandra observations of the Type Ia SNR G344.7--0.1 
revealed the distribution of the SN ejecta. 
The X-ray images and radial profiles of the surface brightness have revealed a centrally peaked 
distribution of the Fe ejecta, surrounded by an arc-like structure of the IMEs.
The centroid energy of the Fe~K emission is lower in the central Fe-rich region than in 
the outer IME-rich regions, suggesting that the Fe ejecta were heated by the reverse shock more recently
in the IME regions. These results are consistent with predictions of standard SN Ia models, 
where the heavier elements are synthesized in the interior of an exploding white dwarf.\\


This work is financially supported by the {\it Chandra} GO Program grant GO8-19052A
and Grants-in-Aid for Scientific Research (KAKENHI) of the Japanese Society for the
Promotion of Science (JSPS) grants No.,\,19H00704 (HY), and 24224005 and 19H05075 (HS).
S.K. is supported by the Leading Initiative for Excellent Young Researchers, Ministry of Education,
Culture, Sports, Science and Technology (MEXT), Japan.
H.S. is supported by ``Building of Consortia for the Development of Human Resources
in Science and Technology" of MEXT (grant No.\,01-M1-0305).
NANTEN2 is an international collaboration of 11 universities: Nagoya University,
Osaka Prefecture University, University of Bonn, University of Cologne,
Seoul National University, University of Chile, University of New South Wales,
Macquarie University, University of Sydney, University of Adelaide, and University of ETH Zurich.


{\software{CIAO \citep{Fruscione06}, XSPEC \citep{Arnaud96}}

\appendix

\section{Background Estimation}
\label{sec:background}

An X-ray background consists of three major components, non X-ray background (NXB), 
cosmic X-ray background (CXB), and the Galactic thermal emission. The third component 
is non-negligible in our study, since G344.7--0.1 is located in the Galactic plane.

\begin{deluxetable}{cCC}[ht!]
\tablecaption{Best-fit Parameters for the Background Model \label{tab:bpar}}
\tablenum{7}
\tablewidth{0pt}
\tablehead{
\colhead{Components} & \colhead{Parameters} & \colhead{}
}
\startdata
NXB & \mathit{\Gamma}_1 & 0.32^{+0.01}_{-0.01}\\
 & \mathit{\Gamma}_2 & -0.78^{+0.04}_{-0.04}\\
 & E_{\mathrm{break}}\ (\mathrm{keV})& 5.32^{+0.11}_{-0.13}\\
 & K_{\mathrm{NXB}} \tablenotemark{c}& 5.68^{+0.06}_{-0.06}\times 10^{-2}\\
 Absorption \tablenotemark{a} & N_{\mathrm{H}}\ (10^{22}\ \mathrm{cm^{-2}})& 7.34^{+0.25}_{-0.26}\\
CXB & \mathit{\Gamma} & 1.40\ (\mathrm{fix})\\
 & K_{\mathrm{CXB}} \tablenotemark{c}& 5.65\times 10^{-5}\ (\mathrm{fix})\\
Galactic thermal emissions \tablenotemark{b} & kT_{\mathrm{apec1}}\ (\mathrm{keV})& 0.35^{+0.02}_{-0.02}\\
 & \text{VEM} \tablenotemark{d}& 0.26^{+0.06}_{-0.05}\\
 & kT_{\mathrm{apec2}}\ (\mathrm{keV}) & 0.79^{+0.17}_{-0.18}\\
 & \text{VEM } \tablenotemark{d}& 4.76^{+8.71}_{-4.28}\times 10^{-3}\\ \hline
 C-statistic/dof & & 1410.75/1152\\
\enddata
\tablenotetext{a}{The absorption cross sections are taken from \cite{Verner96}.}
\tablenotetext{b}{Solar abundances of \cite{Wilms00} are assumed.}
\tablenotetext{c}{In units of $\mathrm{photon\ keV^{-1}\ cm^{-2}\ s^{-1}}$ at 1 keV.}
\tablenotetext{d}{In units of $\mathrm{10^{14}\ cm^{-5}}$.
The volume emission measure (VEM) is given as $\int n_{\mathrm{e}}n_{\mathrm{H}}dV/(4\pi D^2)$,
where $V$ and $D$ are the volume of the emission region ($\mathrm{cm^3}$)
and the distance to the emitting source ($\mathrm{cm}$), respectively.}
\end{deluxetable}

\begin{figure}[ht!]
\gridline{\fig{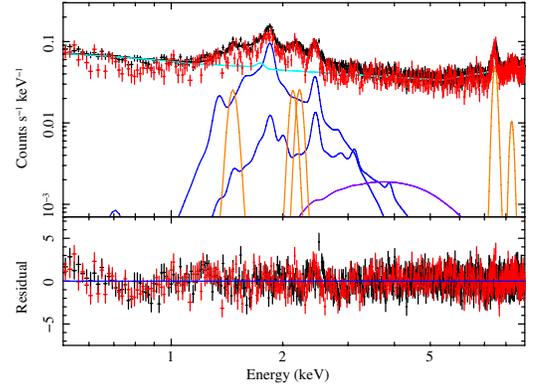}{0.8\columnwidth}{}}
\caption{Raw spectra of the background region around G344.7--0.1.
The black and red data are the data from the May and July observation series, respectively.
The solid lines represent the following models for the spectrum taken in 2018 May,
black: best-fit model; light blue: NXB continuum;
orange: instrumental emission lines; blue: Galactic thermal emissions; and purple: CXB.
\label{fig:bspec}
}
\end{figure}

The NXB spectra contain several fluorescence lines and continuum emission.
We reproduce the NXB component with a model of the {\tt bknpowerlaw} $+$ 5 Gaussians.
Gaussians are for instrumental emission lines of Al K$\alpha$ ($\sim$1.47\,keV),
Au M$\alpha$ ($\sim$2.11\,keV), Au M$\beta$ ($\sim$2.20\,keV),
Ni K$\alpha$ ($\sim$7.47\,keV), and Ni K$\beta$ ($\sim$8.29\,keV).
We assume {\tt tbnew\_gas} $\times$ {\tt powerlaw} model for the CXB emission.
The typical value of CXB normalization is
$N_{\mathrm{CXB}}$\,$\sim$\,8--10\,photon\,keV$^{-1}$\,cm$^{-2}$\,s$^{-1}$\,sr$^{-1}$
at 1 keV \citep[][Suzaku]{Nakashima18}. 
The solid angle of our background region is
$(7^2-5^2)\pi$\,arcmin$^2$\,$\sim 6.28\times 10^{-6}$\,sr,
so we estimate the CXB normalization
$K_{\mathrm{CXB}}=5.65\times 10^{-5}$\,photon\,keV$^{-1}$\,cm$^{-2}$\,s$^{-1}$ at 1 keV.
We fix ${\it \Gamma}=1.40$ and
$K_{\mathrm{CXB}}=5.65\times 10^{-5}$\,photon\,keV$^{-1}$\,cm$^{-2}$\,s$^{-1}$ at 1 keV for the CXB {\tt powerlaw}.
For the final component, Galactic thermal emissions,
which are dominant in the 1--5\,keV band,
we use two absorbed APEC models, as {\tt tbnew\_gas} $\times$ ({\tt apec}+{\tt apec}).
Metal abundances of the APEC components are assumed to be the solar value of \cite{Wilms00}.
We link the absorption column density $N_{\mathrm{H}}$ to that of the CXB.
Table \ref{tab:bpar} gives the best-fit parameters of our background model.
This best-fit model is shown in Figure \ref{fig:bspec}.

The intensities of the Galactic center X-ray emission (GCXE) and Galactic ridge X-ray emission (GRXE)
are modeled by \cite{Uchiyama13} based on many {\it Suzaku} observations.  We compared the intensity
of the Galactic emission component in our model with the predictions of the GCXE and GRXE intensities
in this paper. We calculated the intensity in the 2.3--5.0\,keV band based on equation (1) in \cite{Uchiyama13}
and derived a value of $12.6\times10^{-7}$\,photon\,s$^{-1}$\,cm$^{-2}$\,arcmin$^{-2}$
which compares with the value in our model of
$12.1^{+4.2}_{-9.3}\ \times10^{-7}$\,photon\,s$^{-1}$\,cm$^{-2}$\,arcmin$^{-2}$ (within 90\% errors).
Therefore, our model intensity for the Galactic emission is consistent with the model predictions in \cite{Uchiyama13}.


\section{Radio Observations}
\label{sec:radio}

\subsection{CO and HI}
\label{subsec:co-h1}

Observations of $^{12}$CO ($J$=1--0) line emission at 115.271202\,GHz were executed
in 2012 May 18th using the NANTEN2 millimeter/submillimeter radio telescope.
The telescope is installed at 4865\,m altitude of the Atacama Desert in Chile, operated by Nagoya University.
We observed an area of 1$^{\circ}\times$1$^{\circ}$ region using the on-the-fly mapping mode with Nyquist sampling.
The 4\,K cooled Nb superconductor-insulator-superconductor mixer receiver was used for the front end.
The system temperature including the atmosphere was $\sim$200\,K in the double-side band.
The backend was a digital Fourier-transform spectrometer with 1684 channels or 1\,GHz bandwidth,
corresponding to a velocity coverage of 2600\,km\,s$^{-1}$ and a velocity resolution of 0.16\,km\,s$^{-1}$.
After convolving the cube data with a two-dimensional Gaussian kernel of $90''$,
the final beam size was $\sim 180''$ in FWHM.
The absolute intensity was calibrated by observing the standard source IRAS~16293--2422
[($\alpha_{\rm J2000}$, $\delta_{\rm J2000}$)=(16$^{\rm h}$32$^{\rm m}$23$^{\rm s}$.3, --24$^{\circ}$28$'$39$''$.2)]
\citep{Ridge06}.
We also observed IRC+10216 every day, and the pointing accuracy was better than $10''$.
The  typical noise fluctuation of the final dataset is $\sim$0.65\,K at the velocity resolution of 0.63\,km\,s$^{-1}$.

The HI line data at 1.4\,GHz is from the Southern Galactic Plane Survey \citep[SGPS;][]{McClure05}.
The archival dataset was obtained using a combination of the Australia Telescope
Compact Array (ATCA) and Parkes radio telescope. The angular resolution is $130''$
and the velocity resolution is 0.82\,km\,s$^{-1}$.
The typical noise fluctuation is $\sim$1.9\,K at the velocity resolution of 0.82\,km\,s$^{-1}$.

\subsection{Estimation of Total ISM Protons}
\label{subsec:ISM}

To obtain the total proton column density map, we estimate proton column densities
in both the molecular and atomic clouds \citep[e.g.,][]{Fukui12,Fukui17,Kuriki18,Sano19}.
The proton column density of molecular cloud $N_{\rm p}({\rm H_2})$ can be estimated
from the following equations:
\begin{eqnarray*}
N({\rm H_2})=X_{\rm CO}\cdot W({\rm CO})\ ({\rm cm}^{-2}) \\
N_{\rm p}({\rm H_2})=2\times N({\rm H_2})\ ({\rm cm}^{-2})
\end{eqnarray*}
where $N$(${\rm H_2}$) is the column density of molecular hydrogen in units of cm$^{-2}$,
$X_{\rm CO}$ is the CO-to-H$_2$ conversion factor in units of (K\,km\,s$^{-1}$)$^{-1}$ cm$^{-2}$,
and $W({\rm CO})$ is the integrated intensity of CO in units of K\,km\,s$^{-1}$.
In the present paper, we used $X_{\rm CO}=2.0\times 10^{20}$\,(K\,km\,s$^{-1}$)$^{-1}$ cm$^{-2}$
\citep{Bertsch93}.

On the other hand, an estimation of the proton column density of atomic cloud
$N_{\rm p}({\rm HI})$ is tricky because we should consider the optical depth of HI.
According to \cite{Fukui15}, 91\% of atomic hydrogen in local clouds has
an optical depth of 0.5 or higher with respect to the HI line emission. Therefore we cannot use
the well-known equation assuming the optical depth of HI $\ll$ 1 \citep[e.g.,][]{DL90}:
\begin{eqnarray*}
N_{\rm p}({\rm HI})=1.823\times 10^{18} W({\rm HI})\ ({\rm cm}^{-2})
\end{eqnarray*}
where $W({\rm HI})$ is the integrated intensity of HI in units of K\,km\,s$^{-1}$.

To estimate the optical-depth-corrected proton column density of atomic cloud $N_{\rm p}({\rm HI})$',
we used a relation between $W({\rm HI})$ and $N_{\rm p}({\rm HI})$' presented by \cite{Fukui17}.
The authors derived $N_{\rm p}({\rm HI})$' as a function of $W({\rm HI})$ using the dust opacity map
at 353\,GHz obtained from IRAS and Planck datasets \citep{Planck14}
taking into account nonlinear dust properties \citep[e.g.,][]{Roy13,Okamoto17,Hayashi19a,Hayashi19b}.
In the present paper, we derived a conversion factor from $W({\rm HI})$ to $N_{\rm p}({\rm HI})$'
fitted by the result of \cite{Fukui17} as a function of $W({\rm HI})$.
We obtained averaged value of $N_{\rm p}({\rm HI})$' toward G344.7--0.1 is
$\sim$1.2$\times 10^{21}$\,cm$^{-2}$, which is $\sim$2.3 times higher than
that of $N_{\rm p}({\rm HI})$ assuming the optically thin case.
Finally, we derived the total proton column density map $N_{\rm p}({\rm H_2}+{\rm HI})$
combining $N_{\rm p}({\rm HI})$' and $N_{\rm p}({\rm H_2})$ whose angular resolution is smoothed
to match the FWHM of NANTEN2 beam size of $\sim$180$''$ (see Figure\,\ref{fig:co-h1_map}).

\subsection{An Expanding Gas Motion}
\label{subsec:expansion}

\begin{figure}[ht!]
\gridline{\fig{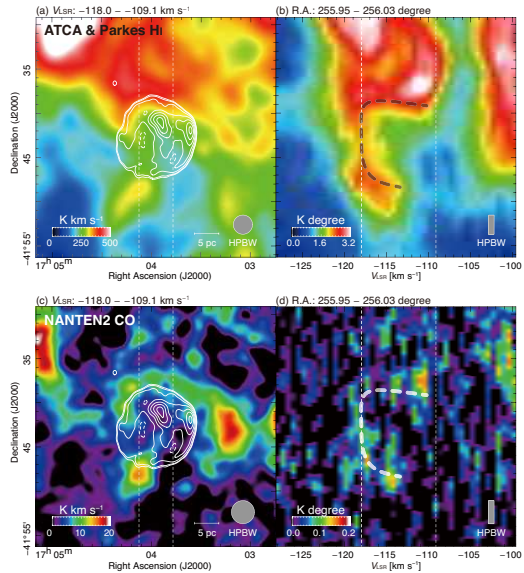}{0.8\columnwidth}{}
	}
\caption{(a), (c) Integrated intensity map of (a) HI and (c) CO
superposed on the MOST 843\,MHz radio continuum.
The integration velocity range and contour levels are the same as in Figure\,\ref{fig:co-h1_map}.
The scale bars at a distance of 6.3\,kpc and half-power beamwidths are also shown
in the bottom right corner. The dashed vertical lines indicate the integration range
of the position-velocity diagram in the Figures\,\ref{fig:co-h1_ind}(b) and \ref{fig:co-h1_ind}(d).
(b), (d) Position-velocity diagrams of (b) HI and (d) CO.
The integration range in R.A. is from 255$^\circ$.95 to 256$^\circ$.03.
The beam size and velocity resolution of each map are shown in the bottom right corner.
The dashed curves indicate a boundary of the CO or HI cavity in the position-velocity diagram (see the text).
\label{fig:co-h1_ind}}
\end{figure}

\cite{Giacani11} argued that G344.7--0.1 is likely associated with an open HI shell
at the central velocity of $\sim$\,115\,km\,s$^{-1}$, corresponding to the kinematic distance of $6.3\pm 0.1$\,kpc.
Here, we investigate the cloud association by using both the CO and HI datasets.

Figures\,\ref{fig:co-h1_ind}(a) and \ref{fig:co-h1_ind}(c) show the integrated intensity maps
of HI and CO at a velocity of $\sim$\,115\,km\,s$^{-1}$, respectively.
We found a presence of shell-like structure not only in the HI map, but also in the CO map.
The CO clouds nicely surround the SNR, especially for the northern half of the shell.
Figures\,\ref{fig:co-h1_ind}(b) and \ref{fig:co-h1_ind}(d) show
the position-velocity diagrams of HI and CO, respectively. We found cavity-like structures
of HI and CO (dashed lines) in the velocity range from --118.0\,km\,s$^{-1}$ to --109.1\,km\,s$^{-1}$,
which have a similar diameter to G344.7--0.1 in terms of the decl.
These cavity-like structures represent an expanding gas motion similar to a ``wind-blown shell,"
originated by strong winds from the progenitor system: e.g., stellar winds from a high-mass progenitor
or accretion winds (also known as ``disk winds") from a single degenerate progenitor system of Type Ia SNR.
For the case of Type Ia SNR G344.7--0.1, the latter origin is favored.
The expansion velocity of molecular and atomic clouds ${\it \Delta}V$ is estimated to be
$\sim$\,4.5\,km\,s$^{-1}$, which is consistent with the previous studies of interstellar environments
in the Type Ia SNRs (e.g., {\it Tycho}, ${\it \Delta}V$$\sim$\,4.5\,km\,s$^{-1}$, \citealt{Zhou16};
N103B, ${\it \Delta}V$$\sim$\,5\,km\,s$^{-1}$, \citealt{Sano18}).
In any case, the SNR shock wave travels through the wind-blown shell during the free expansion phase,
and is finally decelerated when the shockwave encounters the wind wall. Therefore,
the size of the SNR generally coincides with that of the wind-blown shell.
Summarizing the above considerations, both the molecular and atomic clouds in the velocity
range from --118.0\,km\,s$^{-1}$ to --109.1\,km\,s$^{-1}$ are likely associated with the SNR G344.7--0.1,
which is consistent with the previous HI study by \cite{Giacani11}.

\bibliographystyle{aasjournal}
\bibliography{g344_paper_ref}

\end{document}